# Wet Scandium Etching for hard mask formation on a silicon substrate

Julia Bondareva[2], Ekaterina Timofeeva[1], Alexandr Anikanov[1], Maxim Krasilnikov[1], Maxim Shibalov[1], Vasily Sen[1], Alexander Mumlyakov[1], Stanislav Evlashin[2] and Mikhail Tarkhov[1]

[1] Institute of Nanotechnology of Microelectronics of the Russian Academy of Sciences, Leninsky Prospect, 32A, Moscow 119991, Russia

[2] Skolkovo Institute of Science and Technology, 30, bld. 1 Bolshoy Boulevard, Moscow 121205, Russia

E-mail: j.bondareva@skoltech.ru

## Abstract

Nowadays, microelectronics requires the search for new materials, including masks for creating structures. The intermediate hard mask strategy is one of the key issues in achieving a good balance between lithography and etching at the microelectronic fabrication. One of the interesting challenges in microelectronics and photovoltaics is the creation of interspacing, vertically oriented silicon arrays on Si substrate for semiconductor devices with multi-function. The fabrication of such structures is still a serious technological problem and requires searching for new approaches and materials. In this work, we propose using scandium as a new hard mask material over silicon due to its high resistance to plasma chemical etching and low sputtering coefficient. We have shown that a wet etching of the scandium layer with a thickness of several nanometers can be used to obtain pattern structures on silicon with a resolution of up to 4 microns, which is a good result for the wet etching approach. The scandium was found to be an excellent resistant mask over silicon with the lowest etch rate compared to other metal masks under the selected plasma etching conditions. Therefore, a scandium hard mask can open up new possibilities for the formation of different microscale topographical patterns.

## 1. Introduction

Protruded and recessed vertically aligned Si microstructures are of great interest in many fields, such as highly sensitive sensors, [1] [2] solar and thermoelectric materials for energy collection, [3] [4], and batteries. [5] In sensor application, silicon nanowire structures can overcome the limitations of planar chemically-sensitive field-effect transistor, since the one-dimensional morphology and cross-section of nanowires at the nanometer level lead to the accumulation of carriers in the "main mass" of the device, thus providing sufficient sensitivity to detect individual particles. Moreover, in the solar industry, vertical alignment of silicon microwires can separate the light absorption and charge collection directions of minority carriers, and thus can provide excellent performance, provided the microwires are long enough and oriented appropriately. [6] In addition, the use of sub-micrometer and micro Si pillars as electrodes in lithium batteries has shown improvement in battery lifetime. [7] The small diameter of Si pillars allows for better accommodation of the large volume changes without the initiation of fracture that can occur in bulk materials and all Si wires can contribute to capacity through electrical connection to the metal current collector. However, implementing such devices into existing applications will require breakthroughs in size-controlled manufacturing, reduced defect and impurity density, and, most importantly, the ease of manufacturing flexible electronic systems over large areas. Methods for obtaining such structures are divided into bottom-up, for example, chemical vapor deposition (CVD), and top-down, such as lithography, including masking and etching processes. In the lithographic approach, the fabrication of different multilevel structures in microelectronics requires a careful selection of a special combination of masking layers and a specific process sequence.[8] The quality of the final structures can be improved by using extremely selective hard mask materials.

The mask choice is determined by many factors, such as the ability to be deposited, patterned, sustained, and selectively removed. The most common masking material is organic polymers as photoresists; when a photoresist is not resistant to the etchant, a more durable silicon nitride, aluminum oxide, thermal silicon oxide, or chromium and nickel masks are used. [9] The most commonly used rigid masks can be roughly divided into semiconductor oxides, nitrides ($SiO_2$, $Si_3N_4$, etc.), various metal oxides ($Al_2O_3$, $Cr_2O_3$, etc.), and metals (Cr, Ni, etc.). Dielectric materials have required multi-step pattern transfer processes from the pre-fabricated pattern [10] and still suffer from many undercut in smaller via geometries. [11] Metal oxides exhibit high selectivity only at high biases and lead to undercut due to the apparent difference in their electrical properties compared to silicon. [12] Metals are usually formed by the lift-off method, but this can lead to significant pattern distortion, and metal etching can cause metal pollution problems due to high sputtering coefficient values. [13][14] Silicon nitride deposited by low pressure CVD is often used as a hard mask in microfabrication, but due to its instability in fluorine-based plasma [15] its use is limited to wet chemical etching. Wet etching of nitride films is often performed in concentrated hot orthophosphoric acid ($H_3PO_4$) at 150°C to 180°C, this condition is not always suitable for pattern formation and is not always compatible with most masks and chip architectures. Considering several limitations of already existing masks, searching and applying new hard masks with high chemical resistance to plasma-chemical processes of silicon etching is an ongoing challenge. Plasma-chemical etching of silicon is carried out in chlorine or fluorine-containing plasma, for example, $CCl_2F_2/O_2$ and $SF_6/O_2$ mixtures. [16][17] Therefore, a selected masking material should be stable in these conditions.

In this study, we propose scandium as a promising materials for hard mask formation in micro and nanoelectronics. It is well known that metals are very resistant to $SF_6$ - $O_2$ plasma and, in principle, can be ideal masking materials. [18] Metals tend to sputter during the plasma etching process, resulting in uncontrolled re-deposition of metal atoms on the substrate and the formation of needles within the etched structures. [19] This negative phenomenon is not typical for scandium since the sputtering coefficient for scandium is significantly lower than that of nickel and chromium. [20] Moreover, scandium is of high technical and economic interest due to its use in high-tech materials, [21] in electric light sources, aerospace and military, chemical industry, metallurgy, and other fields. [22][23] Despite the high cost, a few hundred thousand dollars per kilogram, [24] scandium can be used as nanometers thick mask material without significantly increasing the cost of the final product.

Besides the selection of masking components, the choice of the etching condition is an important step in the production of structures in microelectronics. The fabrication of structures in bulk silicon requires the development of technological processes with high selectivity in the etching rates of masks and silicon. The first technological solution for silicon etching is the Bosch process, [25][26] but its use requires a special mask, passivation film, and silicon substrate preparation. The second solution is an isotropic plasma-chemical etching of silicon for the formation of high-density structures with silicon emission centers. [27] Scandium possesses high fluorine chemistry ($SF_6$) stability, making it resistant to dry etching with reactive gases. [28] Therefore, wet chemical etching is the most suitable method for forming a pattern on a scandium mask. [29] Wet etching has several advantages: ease of implementation, low cost, good material selectivity, and a fast process due to a high etching rate, which leads to minimal time contact of the protective layer with the etchant and reduces possible destructions. Very often aggressive media are used for wet etching, such as alkalis, acids, acid mixtures, gaseous etching products, which can negatively affect the surface morphology and fine structures with prolonged exposure. Therefore, an increase in etching rates leads to a minimum contact time of the protective layer with the etchant and reduces possible damage. Despite this, wet etching also has some disadvantages, for example, due to the downscaling trend of semiconductor devices, wet etching can cause problems, especially difficult control of the etch rate and blockage due to poor surface wetting.

In this paper, we demonstrate a Si micropillar structure fabrication method by plasma-chemical etching via patterned scandium mask. Taking into account useful properties of scandium such as a low sputtering coefficient and stability in the process of chemical plasma etching, in this work, we have shown for the first time the possibility of using this metal as a hard mask material over silicon etching. Due to the chemical properties of scandium, its etching is carried out by wet chemistry through a layer of silicon dioxide at room temperature. The hard scandium mask enables high pattern transfer fidelity on silicon with a capability to produce structures with orders of magnitude thicker than the original mask thickness.

## 2. Experimental section

### 2.1. Materials and Reagents

To form microstructures, we used a 4-inch n-type silicon substrate, a scandium target (Girmet Ltd) with a purity of 99.95%, and photoresist AZ 4999. Solutions of 31% hydrochloric acid (HCl) by Sigmatec and 69% nitric acid ($HNO_3$) by Reachem were used for wet chemical etching of scandium.

### 2.2. Fabrication of dense paternal structure in bulk silicon

A schematic diagram of the procedures used to fabricate the 3D Si microstructures via the creation of a multilayer structure is shown in Figure 1.

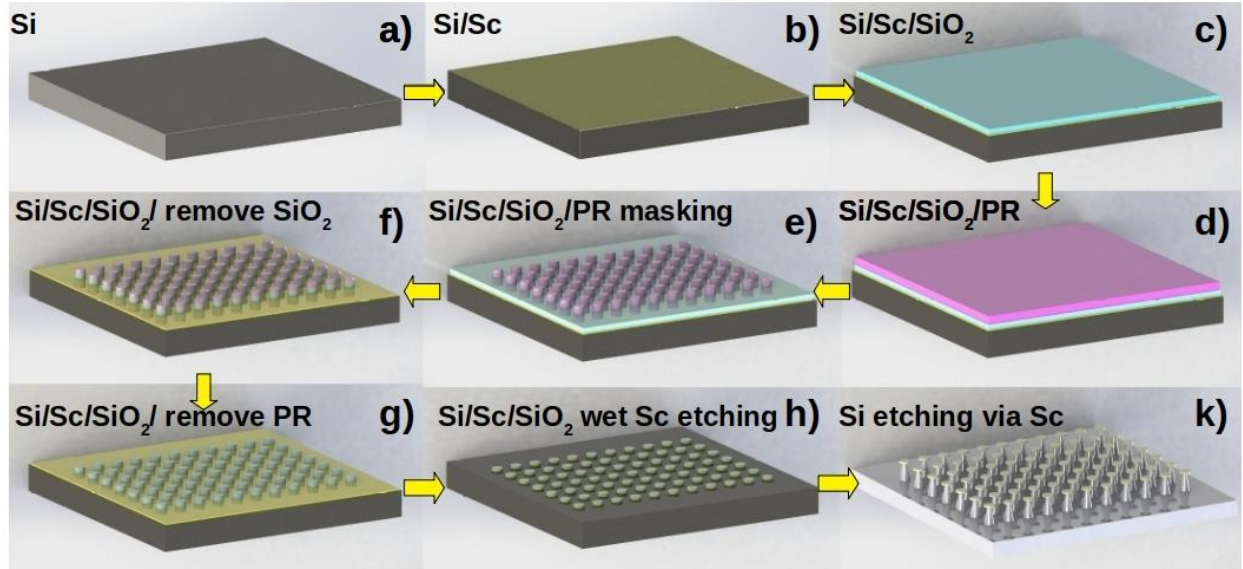

**Figure 1.** Schematic representation of the technological fabrication of a dense paternal structure in bulk silicon in $SF_6$ plasma through a scandium mask. The procedure includes a sequence of steps: a) washing a 4-inch silicon wafer, b) magnetron deposition of scandium, c) PECVD (plasma-enhanced) silicon oxide deposition, d) spin coating photo resistive mask deposition, e) photolithography process, f) plasma-chemical etching of silicon oxide through a resistive mask, g) photoresist removal, h) wet scandium etching, k) plasma-chemical etching of silicon and silicon oxide

Before the scandium deposition process, a silicon substrate with a diameter of 4 inches was pretreated with 1.5 kV ions to activate the surface for bonding **Figure 1a**. Scandium was deposited by the method of magnetron sputtering of a scandium target in an argon atmosphere **Figure 1b**. The argon flow rate was 50 sccm and the magnetron power was 2.5 W/cm$^2$. The deposition rate was 17 nm/min. The thickness of the obtained scandium films was ~ 25 nm.

Silicon oxide, which was formed by the PECVD method from the precursor monosilane ($SiH_4$), was used as a mask for scandium etching **Figure 1c**. The thickness of the obtained silicon oxide was 500 nm. To form a pattern of scandium etching in silicon oxide, exposure was formed according to a given pattern in an AZ 4999 photoresist, 600 nm thick by laser lithography **Figure 1d**, followed by development in an aqueous solution of KOH with a concentration of 0.92% **Figure e**.

Silicon oxide was etched by the plasma-chemical method in an Ar / $SF_6$ gas mixture, and the gas ratio was ~ 2: 3, respectively. The power of the RF source and ICP are 30W and 600W, respectively. The pressure in the process chamber was 8 mTorr **Figure 1f**. Then the photoresist was removed in acetone **Figure 1g**.

After removing the photoresist, liquid etching of scandium was carried out through silicon oxide in various etchants **Figure 1h**. The prepared plates were immersed in the solution of concentrated and diluted $HNO_3$ (2:3) and diluted HCl (1:1) from 40 seconds to 1 minute. Etching occurs in several steps. First, the substrate is oxidized by acids, then the oxidized material from the surface is dissolved into the solution through the formation of chemical substances. Finally, the dissolved species diffuse into the bulk of the solution allowing the surface to contact with fresh etchant solution. Such slow diffusion can happen due to convection induced by thermal gradients, exothermic reactions, gas bubble formation, and mechanical agitation. [30] After wet etching, plates were rinsed with distilled water and washed with acetone and ethanol in an ultrasonic bath for 2 minutes.

The last process to investigate the possibility of using scandium as a hard mask for bulk silicon etching was the process of plasma-chemical etching of silicon **Figure 1k**. The pressure in the process chamber was 5 mTorr in an $SF_6$ gas atmosphere at the power of the RF and ICP sources equal to 5 W and 1000 W, respectively. The stage temperature was 50 °C. The gas flow rate was 30 sccm. The etching rate of silicon and silicon oxide was 1 μm/min and 0.19 μm/min, respectively.

## 2.3. Microstructure Characterization

The morphology and fine structure of multilayer samples were studied using Scanning Electron Microscopy (SEM) Thermo Fisher Scientific in Apreo system with an accelerating voltage of 5 kV (probe current: 50 pA, detector: ETD, tilt angle: 52°). An energy dispersive X-ray analysis (EDX) was used to identify the elemental composition in specific positions. Focused ion beam (FIB) was used to cut the samples to analyze the cross-section beneath the surface. The cross-section was imaged by SEM, and EDX was conducted along the depth direction.

## 2.4. X-ray reflectometry

The thickness of the scandium layer before the etching process as well as samples after etching in $SF_6$ for 5, and 10 minutes were determined by X-ray reflectometry (XRR). The measurements were carried out on an X-ray diffractometer Panalitycal Empyrean, using a proportional detector. Reflectograms were recorded in ω-2θ scanning mode using Cu K α radiation ($\lambda = 1.540605$ Å) at 45 kV and 40 mA. The optic scheme includes a primary beam (parallel mirror, divergence slit 1/32, attenuator with attenuation factor 2.4), diffracted beam (parallel plate collimator, slit 0.18°, programmable attenuator). An anti-scattering knife for reflectometry was installed over the sample during the survey. The reflectograms were processed using the AMASS software; the fitting curves were built using the Powell method.

## 3. Result and Discussion

The technological process of 3D Si microstructure fabrication procedures consists of two main processes, including wet and plasma chemical etching. The first is patterning and etching the scandium metal 25 nm thick film on the Si substrate through the $SiO_2$ masking layer via wet chemical etching. The second is etching the Si substrate using plasma-chemical etching to increase the height profile of the patterned Si structures. Scandium wet etching was chosen because scandium is stable under fluorine chemistry in the plasma etching process and can only be removed by acids, which do not react with the $SiO_2$ masking layer. Thus, wet etching of scandium makes it possible to create a Sc masking layer to further form patterned 3D Si microstructures by a plasma etching process. Moreover, the stability of scandium to reactive gases allows the use of even thin nm-scale films, which is economically justified from the high scandium cost.

### 3.1. Dry silicon etching with scandium mask

To evaluate the selectivity of the scandium mask to the silicon substrate, a typical fluorine-based Si-etch was performed for 5 and 10 min. We used plasma generated from $SF_6$ because F chemically etches Si at a very high rate. However, the rapid chemical etching of Si by F can lead to significant lateral etching. To reduce lateral etching, $O_2$ is usually added to the $SF_6$ plasma to form a thin passivation layer of fluorinated silicon oxide ($SiO_xF_y$). The thickness of scandium mask before and after each etching step was determined by XRR. The measured and modeled XRR patterns of as-deposited Sc/Si sample and after 5 and 10 min etching are shown in Figure 2. In the as-deposited state, we see clear oscillations as predicted by the Powell model, and we can assume that there are no damages between the Sc and Si layers. Moreover, we measured the XRR patterns of Sc/Si after etching the structures in $SF_6$ plasma at 5 and 10 min. The fringes were observed for all samples were consistent with the model, which indicates the stability of the Sc/Si interface. We see a difference in the XRR oscillation frequency after 5 min etching the structure, and this difference is retained for all samples after etching. These changes indicate the appearance of additional layers. For example, scandium behavior in fluorine plasma can be characterized by the formation of an intermediate layer of scandium sulfide and a surface layer of scandium fluoride. The formation of a scandium fluoride passivation layer prevents further scandium etching.

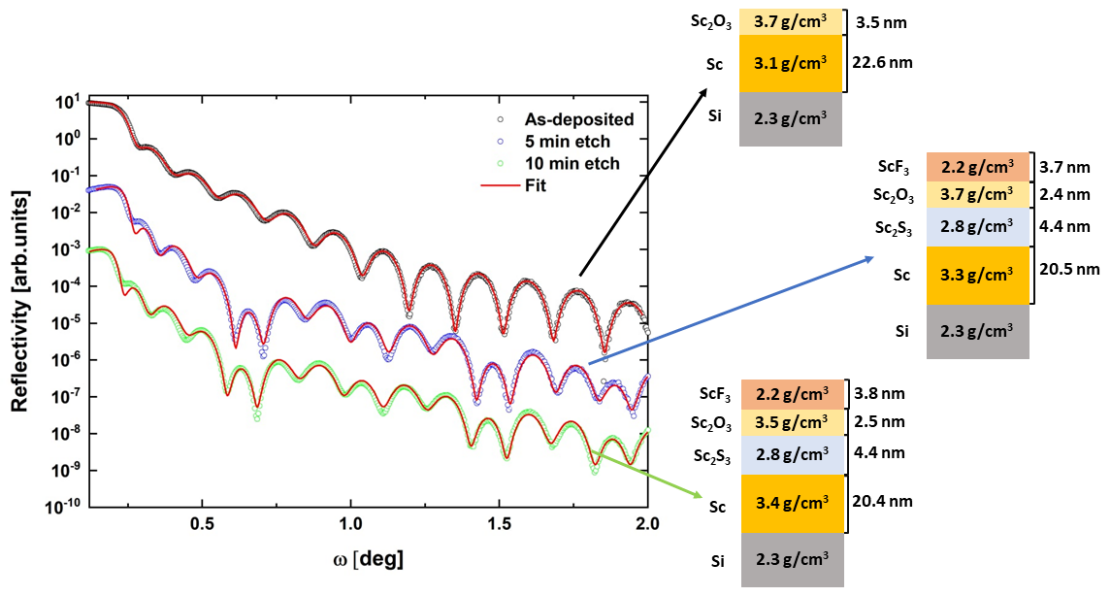

**Figure 2.** Experimental and simulated reflectograms of Sc as-deposited on silicon before etching, after 5 and 10 min of silicon etching. The layer structure is also shown. The error of thickness determination is 0.4 nm

The starting thickness of scandium layer was 22.6 nm and after 5 min of etching the thickness reduced to 20.5 nm and stayed constant after 10 min of etching. The determined etching rate of silicon in our conditions was 1 μm/min. The etching rate of the scandium mask was 0.2 nm/min, which is significantly lower than that of other metal masks, as one can see in Table 1. Selectivity Sc:Si is 1:5000.

**Table 1.** Comparison of metal etch rates in plasma-chemical etching $SF_6+O_2$ [31]

| Metals | Deposition | Etch rates (nm/min) |
|---|---|---|
| Scandium | Sputtered | 0.2 |
| Aluminum | Evaporated | <2.8 |
| Niobium | Ion-Milled | 26 |
| Tantalum | Ion-Milled | 37 |
| Chromium | Evaporated | <1 |
| Chromium | Ion-Milled | <0.9 |
| Molybdenum | Evaporated | 130 |
| Nickel | Evaporated | 0.71 |
| Palladium | Evaporated | 3.1 |
| Platinum | Evaporated | 7.4 |
| TiW (Titanium-Tungsten) (10:90) | Ion-Milled | 550 |
| NiCr (Nickel-Chromium) (80:20) | Evaporated | 3.7 |

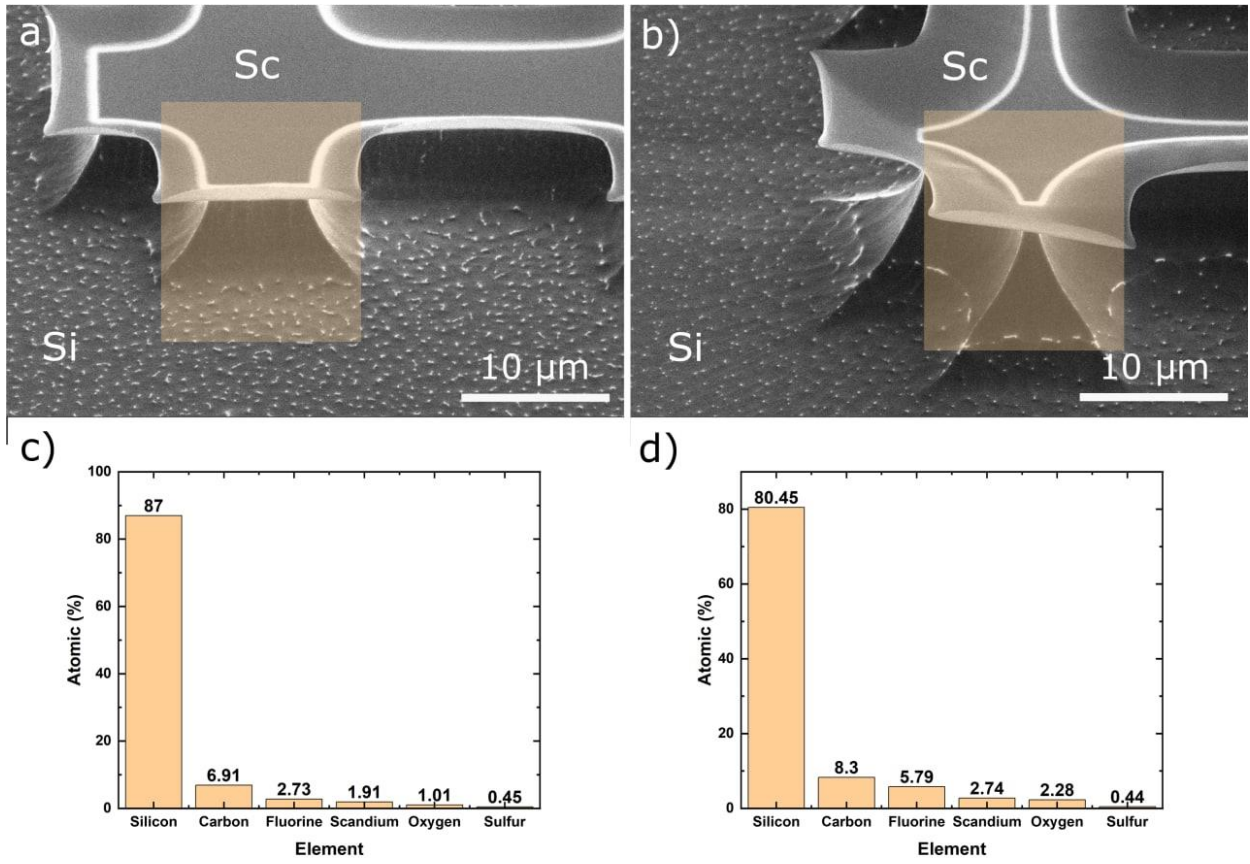

**Figure 3.** SEM cross-section of the microstructures and the EDS analysis. Images correspond to the step shown in Figure 1k. a) SEM image of the microstructure with 5 min silicon etching. b) SEM image of the microstructure with 10 min silicon etching. c) Elemental composition after 5 min silicon etching. d) Elemental composition after 10 min silicon etching. EDS analysis regions indicated in colored squares

The SEM cross-sections in Figures 3 a, b show that the mask undercut and sidewall slope depend on the plasma parameters. In our case, when the $SF_6$ to $O_2$ ratio is high, the mask trim is very strong because the passivation of the sidewall by the O atoms does not keep up with the chemical etching by the F atoms. Moreover, there are defects on the surface of silicon, which can be associated with micro-masking from the Sc etch mask [32] and adsorption of oxygen on the surface, and as a result the appearance of micro-grass in the large open areas on the bottom [33] as you can see in Figure 3. In accordance with the EDX data obtained, the amounts of oxygen and fluorine in the sample after 10 min of etching are increased compared to the sample after 5 min of etching, which confirms the formation of a passivating layer of fluorinated silicon oxide. However, the sulfur atomic ratio is less than expected, which we believe may be due to diffusion of $S_2$ gasses due to in-situ annealing in SEM-EDS imaging. The following article [34] also reports on a similar phenomenon. Also, due to the small thickness of the resulting scandium sulfide (less than 1 µm) according to XRR data, it can be translucent for electrons and a significant part of the received signal will come from the substrate, reducing the calculated sulfur percent.

## 3.2. Wet scandium etching

Particular attention should be paid to wet chemical etching because it helps create a pattern structure for Si etching. Wet etching is an isotropic process that can produce microchannels with undesirable rounded sidewalls. The shape and angle of these microchannels may be adjusted and reduced by applying different parameters during wet etching, such as etchant concentration, exposure time, and usage of a receding mask. [35] [36]

When selecting etchants for our process, their sensitivity to scandium, and their inertness to silicon oxide must be taken into account. Also the possibility of using dilute acid solutions is preferable to control the etching process. In addition, the resistance of scandium to reactive gases makes it possible to use even thin nanometer-sized films as a mask for further silicon etching, which is also economically viable from the point of view of the high scandium cost. The wet-etching process can reduce the surface damage, which the dry-etching process may cause. [37] Some disadvantages of wet etching should be kept in mind; for example, etching is very isotropic, and the acid will affect the substrate in all directions. In this case, the method of anisotropic etching is preferable if applied to the structures used, as mentioned in the works. [38][39] Also, the chemicals need to be changed regularly to maintain the same initial etch rate, and the post-etch treatment should effectively remove any contaminants. Before proceeding to scandium etching, we carefully analyzed the literature on the chemical properties of this material. According to reference [29], nitric, hydrochloric acids, and aqua regia ($HNO_3$:3HCl) are the main etchants of this metal, so below, we considered chemical reactions for all three cases. Chemical etching of scandium, despite its simplicity, was a rather time-consuming process, given the lack of ready-made protocols for such objects, so we decided to optimize this stage by checking different etching modes and analyzing their effect on the final structure.

Scandium is a soft transitional element with a silver-white tint; upon contact with air, it acquires a slightly yellowish or pinkish tint due to scandium oxide ($Sc_2O_3$) formation on the surface. [40] It reacts with water when heated to form hydrogen gas. Scandium readily oxidizes in air, react with acid vapors, and dissolves in most dilute acids; for example, hydrochloric, nitric acids and their mixture.

### 3.2.1 Etching by hydrochloric acid

Scandium easily dissolves in dilute hydrochloric acid to form solutions containing the $Sc^{+3}$ ions together with gaseous hydrogen:

$$2Sc + 6HCl \rightarrow 2Sc^{3+} + 6Cl^- + 3H_2$$

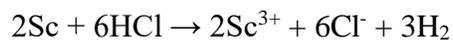

However, the formation of hydrogen gas in the etching reaction is challenging for a homogeneous etching result. The constantly produced $H_2$ bubbles stick to the surface and block or slow down the etching process. In this case, it can help interrupt the etching process by dipping in water, which temporarily removes the $H_2$ bubbles. In order to adjust the scandium wet etching technology, structures with various patterns were fabricated, including both columns and holes, as shown in Figure 4 a and b, respectively. According to our observations, scandium etching even occurs under the influence of vapors of hydrochloric and nitric acids, as shown in Figure 4 for both patterned structures. The samples were stored under acid vapors in a fume hood for several seconds. Scandium etching is not complete, and the surface morphology has an island-like shape with scandium residue.

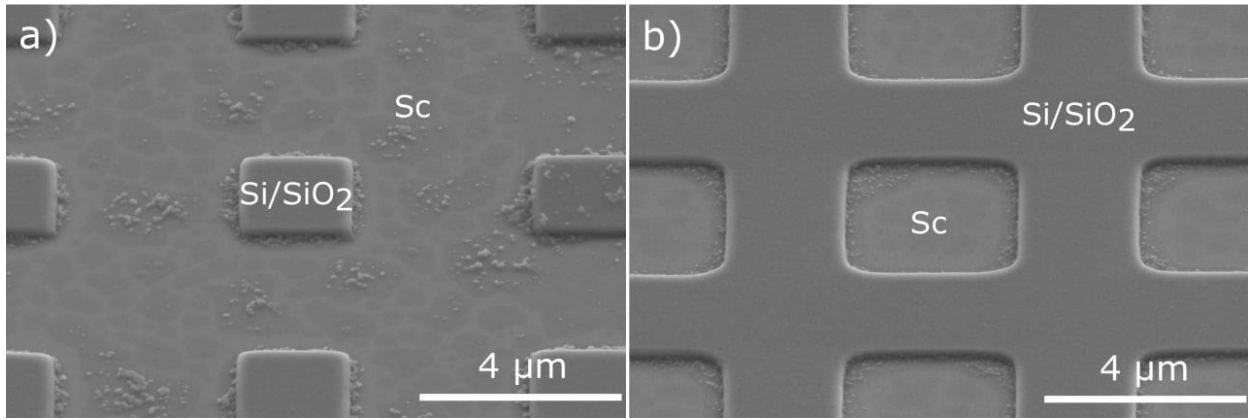

**Figure 4.** SEM images of scandium partial dissolution under nitric and hydrochloric acid vapors. a) 3x2 μm designed size columns, 284 nm height b) 3x2 μm designed size holes, 218 nm height. Images represent different patterned structures corresponding to the step shown in Figure 1h

In our case, dilute hydrochloric acid as an etchant makes it possible to effectively remove the scandium layer through $SiO_2$ layer and save the fine structure of the sample after silicon plasma etching. At the same time, as can be seen from Figure 5, reducing the exposure time leads to fewer artifacts on the surface than in Figure 6.

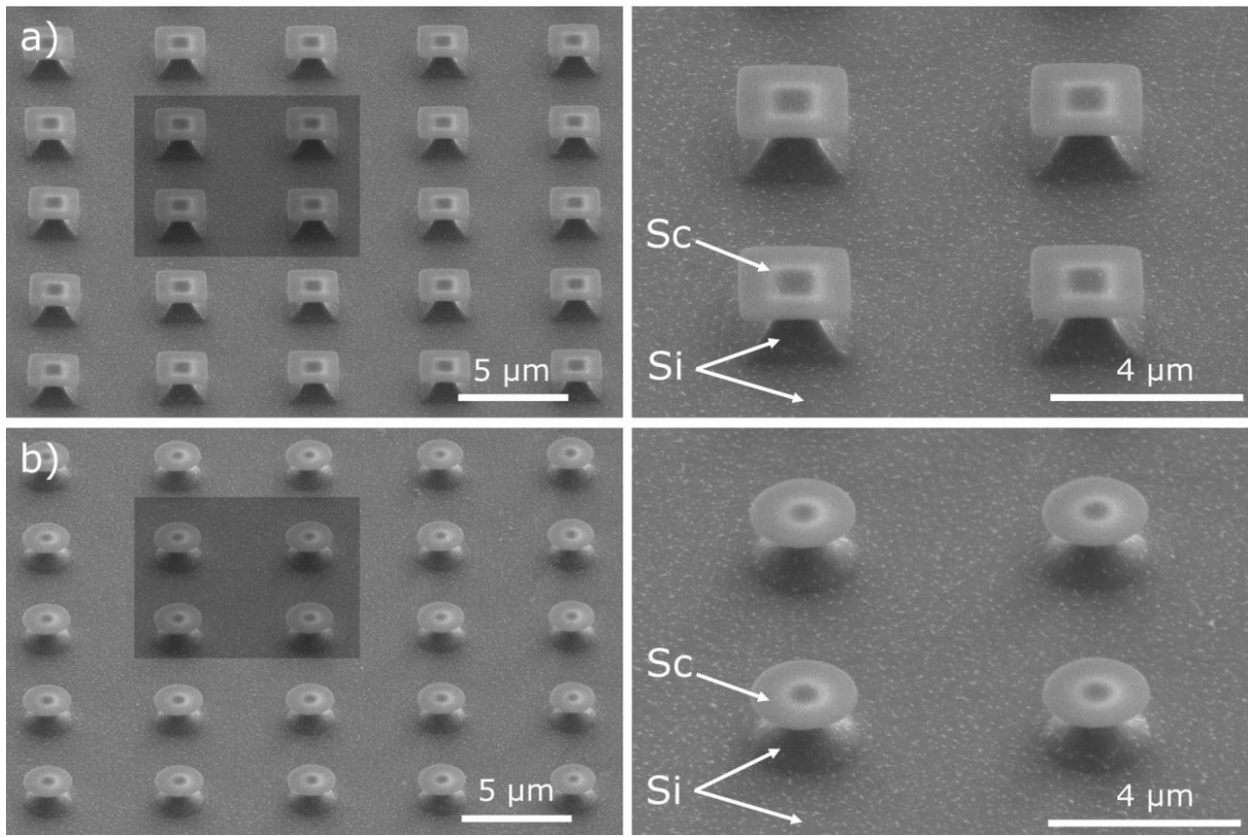

**Figure 5.** SEM images of 1 min silicon etching through scandium mask, which was removed by diluted hydrochloric acid (1:1) for 50 s. The images show different pattern structures with circle (designed d=2 μm) and square-shaped (2x2 μm designed size) vertically oriented silicon arrays with height are 1.354 and 1.438 μm, respectively. Aspect

ratio for the squire and round-shaped structures equal 0.77 and 0.79, respectively. Images correspond to the step shown in Figure 1k

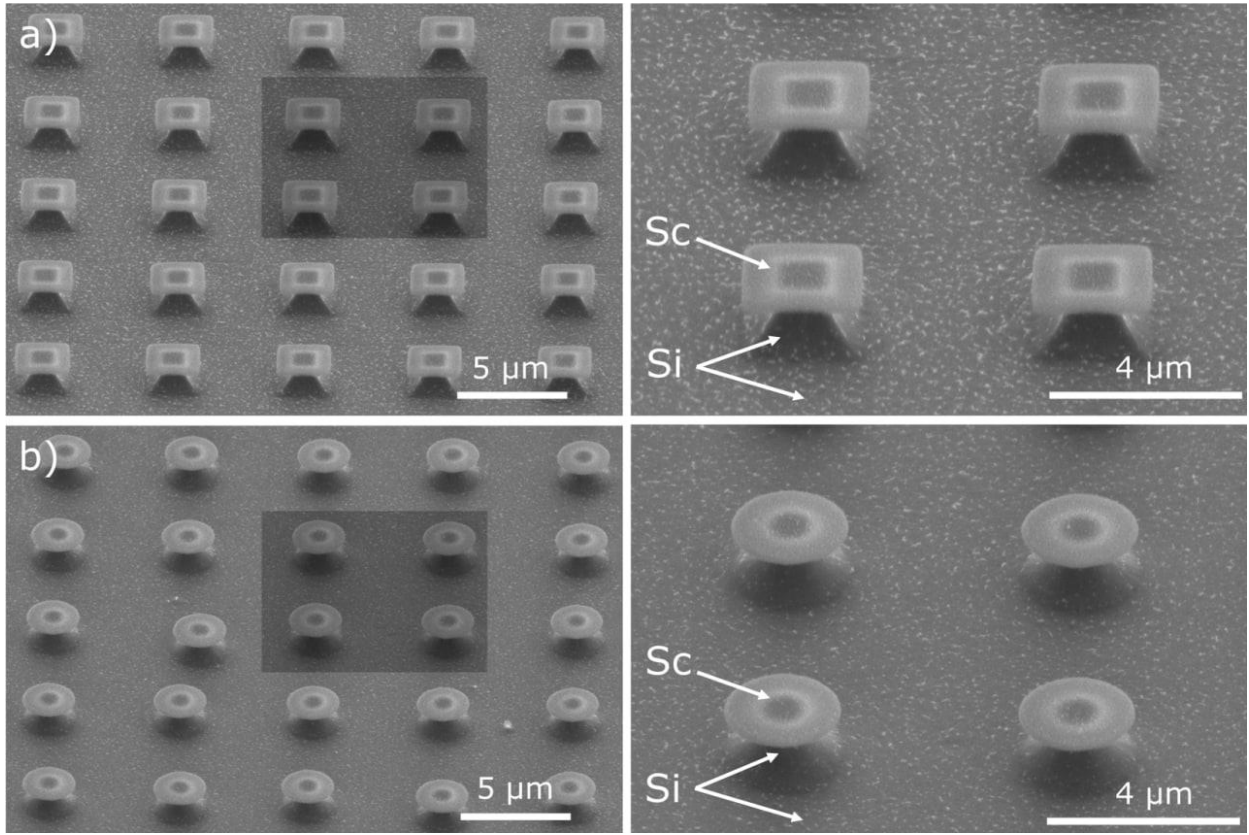

**Figure 6.** SEM images of 1.5 min silicon etching through scandium mask, which was removed by diluted hydrochloric acid (1:1) for 1 min. The images show different pattern structures with circle (designed d=2 μm) and square-shaped (2x2 μm designed size) vertically oriented silicon arrays with height are 1.805 and 1.836 μm, respectively. Images correspond to the step shown in Figure 1k

### 3.2.2 Etching by nitric acid

Scandium can be effectively etched at room temperature in concentrated and dilute nitric acid to produce scandium (III) nitrate, excess of nitric acid can be removed by repeatedly rinsing with distilled water:

$$8Sc + 30HNO_3 \rightarrow 8Sc(NO_3)_3 + 3NH_4NO_3 + 9H_2O$$

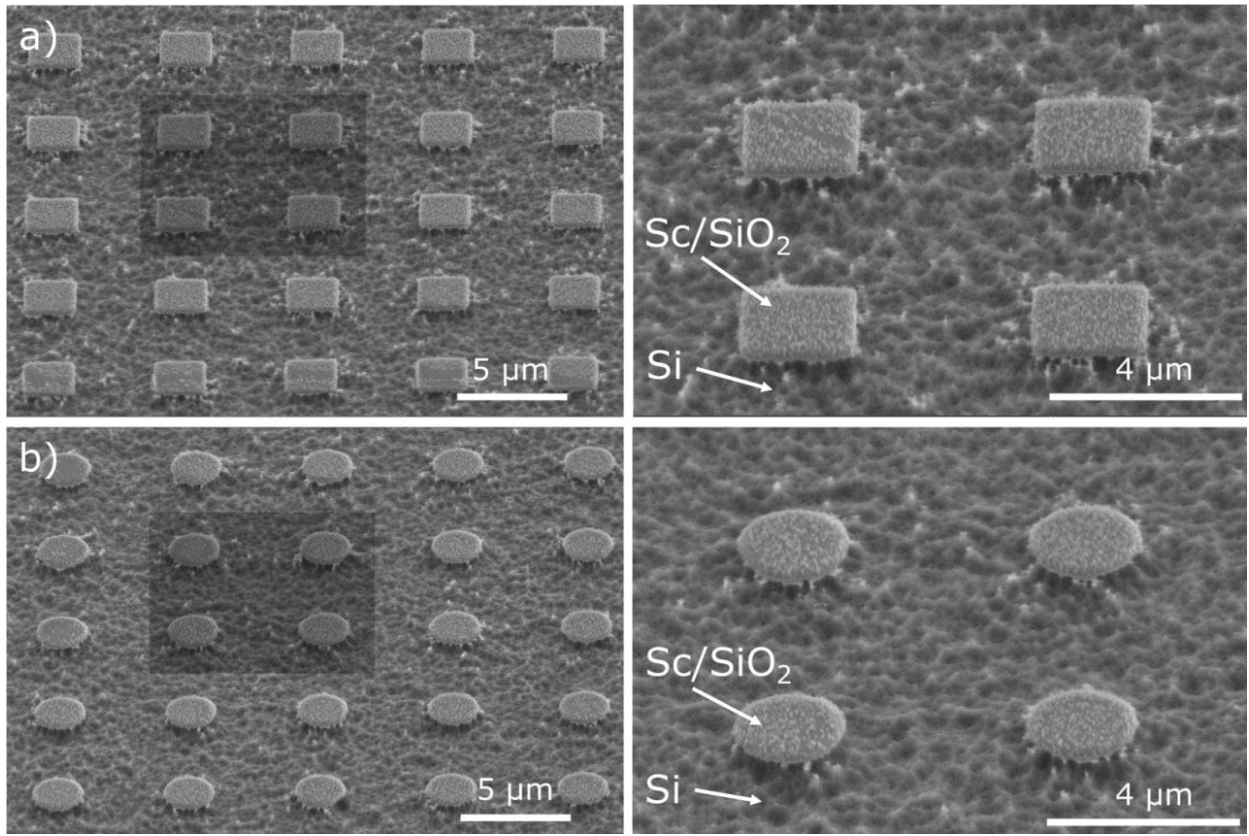

**Figure 7.** SEM images of 0.5 min silicon etching through scandium mask, which was removed by concentrated nitric acid for 40 s. The images show different pattern structures with circle (designed d=2 μm) and square-shaped (2x2 μm designed size) vertically oriented silicon arrays. Images correspond to the step shown in Figure 1k

When using concentrated nitric acid as an etchant, it becomes difficult to control the etching rate. At short acid exposure times, under-etching is observed with subsequent masking of scandium residues on the surface, as shown in Figure 7. Moreover, because of the short etching time of silicon, a roughness is formed on the scandium surface due to the presence of silicon oxide residues. An increasing acid exposure time can lead to the destruction of fine structures and the appearance of defects.

However, scandium does not react with some concentrated acids; for example, scandium is stable with a 1:1 mixture of 69% nitric acid ($HNO_3$) and 48% hydrofluoric acid (HF), possibly, due to the formation of a passive, protective trifluoride layer ($SiF_3$) that prevents further reaction:

$$2Sc + 6HF + 2HNO_3 \rightarrow 2ScF_3 + 2NO + 4H_2O$$

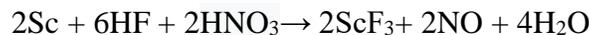

Even though scandium can be easily removed with some concentrated acids, it is preferable to use dilute acids in etching tasks to remove a scandium layer gradually with control of the etching rate.

Better results are achieved when using dilute nitric acid at the same exposure time Figure 8. Dilution will reduce the etching rate.

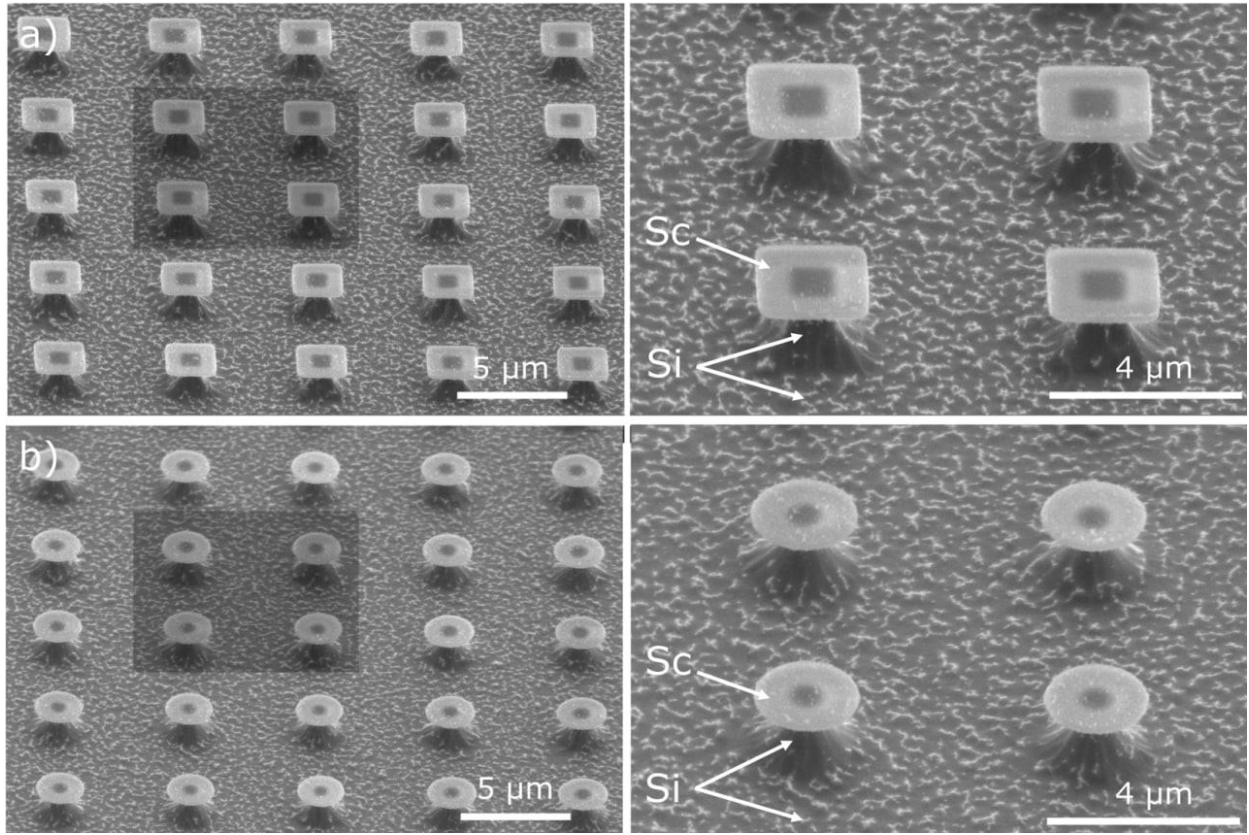

**Figure 8.** SEM images of 1.5 min silicon etching through scandium mask, which was removed by dilute nitric acid (2:3) for 40 s. The images show different pattern structures with circle (designed d=2 μm) and square-shaped (2x2 μm designed size) vertically oriented silicon arrays with height are 1.784 and 1.826 μm, respectively. Images correspond to the step shown in Figure 1k

### 3.2.3 Etching by aqua regia

The mixture of hydrochloric acid and nitric acid in a mixing ratio of 1:3 (AQREG) is the main etching agent of scandium. The very strong oxidative effect of this mixture relates to the formation of nitrosyl chloride (NOCl) via the reaction:

$$HNO_3 + 3HCl \rightarrow NOCl + 2Cl^- + 2H_2O$$

while free Cl radicals formed in the solution keep scandium dissolved as Cl-complex (HScCl$_4$):

$$Sc + HNO_3 + 4HCl \rightarrow HScCl_4 + NO + 2H_2O$$

Taking into account the object of our study (nm-thick scandium film), we analyzed and tried different etching conditions and noticed that for our case, only nitric or hydrochloric acid is enough to achieve the desired structure. In this regard, we don't see any reason to use an acid mixture.

Unremoved scandium can remain after acid etching due to inefficient process parameters, such as non-uniform and short-term exposure and high etchant concentration. It can be seen that in the case of using nitric acid as an etchant, the surface is more contaminated compared to using hydrochloric acid. We have not studied the nature of this residual contamination in detail. Still, during the etching reaction, scandium nitrate and chloride are formed as by-products, which in the air tend to form scandium nitrate, and chloride hydrates $Sc(NO_3)_3 \cdot 4H_2O$ and $ScCl_3 \cdot 3H_2O$, $ScCl_3 \cdot 6H_2O$, respectively. [41] Thus, the residual contamination is most likely formed by crystalline hydrates. In contrast to scandium chloride hydrates, scandium nitrate hydrate is moderately soluble in water, which can explain the differences in surface morphology.

Figures 5, 6, and 8 show that after etching bulk silicon through a thin scandium film (~ 25 nm thick), the mask has no mechanical distortion due to internal stresses. This effect is very important in smooth etching profiles formation of functional structures with an isotropic etching profile.

Nevertheless, similar defects are observed on the surface at different modes of wet scandium etching and at different times of silicon plasma etching. We believe that such defects are caused by different micro-masking present on the silicon surface after plasma etching. Predicting the source of the micro-masking is difficult, as external and internal factors are possible. External sources of these micro-masking vary, such as native oxide or dust on the surface before etching. [42] But micro-masking can also occur in the vacuum chamber during the etching process itself. This internal micro-masking can be caused either by re-deposition of the scandium masking material, atomized by incoming high-energy ions, or by residual surface passivation, or by particles emanating from the chamber walls. [43] In addition, when analyzing possible sources of micro-masking, the aspect ratio of the resulting structures should not be forgotten. Silicon micro-masking is often more pronounced in large open areas with low aspect ratios than in narrow areas with high aspect ratios. [44] This effect is particularly noticeable when the micro-masking is due to surface passivation. The thickness of the passivation layer plays an important role here. In our case, in areas with a large spacing more than 5 μm, with low aspect ratio, the structures will receive more oxygen, which can result in active growth of the passivation layer. Thus, in the subsequent removal step, parts of the passivation film may remain on the surface, causing the micro-mask to be under-removed from the surface. This can also be one of the causes of defects on the silicon surface during etching.

## 4. Conclusion

We have shown for the first time that a thin scandium film can be used as a hard mask to create well-defined silicon micropillar structures. It was demonstrated that scandium withstands the plasma-chemical processes used in the silicon etching process and, compared to other commonly used metal masks, has a very low etching rate in fluorine-based plasma. At the same time, wet scandium etching can be carried out in various acids and their mixtures without additional heating. It was shown that the use of diluted hydrochloric acid for scandium etching through the silicon oxide layer is the most optimal condition due to the minimal contamination of by-products. Also, the possibility of silicon plasma etching through thin scandium films without scandium deformation has been experimentally demonstrated. Visualization and analysis of the obtained silicon structures were performed using scanning electron microscopy; the rate of scandium plasma chemical etching was determined by measuring the thickness of the scandium layer by XRR analysis.

The obtained results are of fundamental materials research and open up prospects for use and further investigation of a new hard mask materials for fabrication pattern structures in microelectronics. Moreover, the application of scandium layers with a thickness of no more than 50 nm as a hard mask can be technologically applicable for the lift-off method [45] with a photoresist thickness of 500 nm. We believe that scandium films can be suitable for anisotropic etching of silicon - the Bosch process since this material is highly resistant to fluorine-containing plasma. The possibility of obtaining silicon nanopillar structures using the scandium mask is still controversial and requires additional research.

## Acknowledgements

The authors are grateful to Grants № 0004-2019-0002, № 0004-2019-0003 in support of the Ministry of Science. Fabrication and characterization were carried out at large scale facility complex for heterogeneous integration technologies and silicon + carbon nanotechnologies. We thank Shibalova Anastasia for conducting XRR measurements and Nekludova Polina for FIB and SEM data.

## Data availability statement

All data that support the findings of this study are included within the article.

## References

[1] C. Yi, W. Qingqiao, P. Hongkun, L.C. M., Nanowire Nanosensors for Highly Sensitive and Selective Detection of Biological and Chemical Species, Science (80-. ). 293 (2001) 1289–1292. https://doi.org/10.1126/science.1062711.

[2] F. Patolsky, G. Zheng, C.M. Lieber, Fabrication of silicon nanowire devices for ultrasensitive, label-free, real-time detection of biological and chemical species, Nat. Protoc. 1 (2006) 1711–1724. https://doi.org/10.1038/nprot.2006.227.

[3] M.D. Kelzenberg, S.W. Boettcher, J.A. Petykiewicz, D.B. Turner-Evans, M.C. Putnam, E.L. Warren, J.M. Spurgeon, R.M. Briggs, N.S. Lewis, H.A. Atwater, Enhanced absorption and carrier collection in Si wire arrays for photovoltaic applications, Nat. Mater. 9 (2010) 239–244. https://doi.org/10.1038/nmat2635.

[4] A.I. Hochbaum, R. Chen, R.D. Delgado, W. Liang, E.C. Garnett, M. Najarian, A. Majumdar, P. Yang, Enhanced thermoelectric performance of rough silicon nanowires, Nature. 451 (2008) 163–167. https://doi.org/10.1038/nature06381.

[5] C.K. Chan, H. Peng, G. Liu, K. McIlwrath, X.F. Zhang, R.A. Huggins, Y. Cui, High-performance lithium battery anodes using silicon nanowires, Nat. Nanotechnol. 3 (2008) 31–35. https://doi.org/10.1038/nnano.2007.411.

[6] L. Tsakalakos, J. Balch, J. Fronheiser, B.A. Korevaar, O. Sulima, J. Rand, Silicon nanowire solar cells, Appl. Phys. Lett. 91 (2007) 233117. https://doi.org/10.1063/1.2821113.

[7] K.V. Mironovich, S.A. Evlashin, S.A. Bocharova, M.S. Yerdauletov, S.A. Dagesyan, A.V. Egorov, N.V. Suetin, D.M. Itkis, V.A. Krivchenko, Gaining cycling stability of Si- and Ge-based negative Li-ion high areal capacity electrodes by using carbon nanowall scaffolds, J. Mater. Chem. A. 5 (2017) 18095–18100. https://doi.org/10.1039/c7ta03509h.

[8] B. Bahreyni, Chapter 2 - Microfabrication, in: B.B.T.-F. and D. of R.M. Bahreyni (Ed.), Micro Nano Technol., William Andrew Publishing, Norwich, NY, 2009: pp. 9–46. https://doi.org/https://doi.org/10.1016/B978-081551577-7.50006-7.

[9] R.R. Tummala, M.R. Haley, G. Czornyj, Materials in microelectronics, Ceram. Int. 19 (1993) 191–210. https://doi.org/https://doi.org/10.1016/0272-8842(93)90040-X.

[10] Q. Fang, X. Li, A.P. Tuan, J. Perumal, D.-P. Kim, Direct pattern transfer using an inorganic polymer-derived silicate etch mask, J. Mater. Chem. 21 (2011) 4657–4662. https://doi.org/10.1039/C0JM03869E.

[11] S. Choi, S.J. Hong, Use of Hard Mask for Finer (<10 μm) Through Silicon Vias (TSVs) Etching, Trans. Electr. Electron. Mater. 16 (2015) 312–316. https://doi.org/10.4313/TEEM.2015.16.6.312.

[12] K. Grigoras, L. Sainiemi, J. Tiilikainen, A. Säynätjoki, V.-M. Airaksinen, S. Franssila, Application of ultra-thin aluminum oxide etch mask made by atomic layer deposition technique, J. Phys. Conf. Ser. 61 (2007) 369–373. https://doi.org/10.1088/1742-6596/61/1/074.

[13] K. Lim, S. Gupta, C. Ropp, E. Waks, Development of metal etch mask by single layer lift-off for silicon nitride photonic crystals, Microelectron. Eng. 88 (2011) 994–998. https://doi.org/https://doi.org/10.1016/j.mee.2010.12.113.

[14] H.-Y. Hsieh, S.-H. Huang, K.-F. Liao, S.-K. Su, C.-H. Lai, L.-J. Chen, High-density ordered triangular Si nanopillars with sharp tips and varied slopes: one-step fabrication and excellent field emission properties, Nanotechnology. 18 (2007) 505305. https://doi.org/10.1088/0957-4484/18/50/505305.

[15] C. Reyes-Betanzo, S.A. Moshkalyov, A.C.S. Ramos, J.W. Swart, Mechanisms of silicon nitride etching by electron cyclotron resonance plasmas using SF6- and NF3-based gas mixtures, J. Vac. Sci. Technol. A. 22 (2004) 1513–1518. https://doi.org/10.1116/1.1701858.

[16] L. Louriki, P. Staffeld, A. Kaelberer, T. Otto, Silicon Sacrificial Layer Technology for the Production of 3D MEMS (EPyC Process), Proc. . 1 (2017). https://doi.org/10.3390/proceedings1040295.

[17] L. Wang, M. Liu, J. Zhao, J. Zhao, Y. Zhu, J. Yang, F. Yang, Batch Fabrication of Silicon Nanometer Tip Using Isotropic Inductively Coupled Plasma Etching, Micromachines . 11 (2020). https://doi.org/10.3390/mi11070638.

[18] R. Huber, J. Conrad, L. Schmitt, K. Hecker, J. Scheurer, M. Weber, Fabrication of multilevel silicon structures by anisotropic deep silicon etching, Microelectron. Eng. 67–68 (2003) 410–416. https://doi.org/https://doi.org/10.1016/S0167-9317(03)00097-2.

[19] M. Nie, K. Sun, D.D. Meng, Formation of metal nanoparticles by short-distance sputter deposition in a reactive ion etching chamber, J. Appl. Phys. 106 (2009) 54314. https://doi.org/10.1063/1.3211326.

[20] K. Kanaya, K. Hojou, K. Koga, K. Toki, Consistent Theory of Sputtering of Solid Targets by Ion Bombardment Using Power Potential Law, Jpn. J. Appl. Phys. 12 (1973) 1297–1306. https://doi.org/10.1143/jjap.12.1297.

[21] M. Ochsenkuehn-Petropoulou, L.-A. Tsakanika, T. Lymperopoulou, K.-M. Ochsenkuehn, K. Hatzilyberis, P. Georgiou, C. Stergiopoulos, O. Serifi, F. Tsopelas, Efficiency of Sulfuric Acid on Selective Scandium Leachability from Bauxite Residue, Met. . 8 (2018). https://doi.org/10.3390/met8110915.

[22] W.-H. Yong, W.-Q. Gong, J.-H. Xiao, Y.-S. Zhang, Occurrence State of Scandium in Rare Earth at Yingjiang, Yunnan, in: Mater. Sci. Eng., WORLD SCIENTIFIC, 2017: pp. 656–666. https://doi.org/doi:10.1142/9789813226517_0094.

[23] F. Meng, X. Li, P. Wang, F. Yang, D. Liang, F. Gao, C. He, Y. Wei, Recovery of Scandium from Bauxite Residue by Selective Sulfation Roasting with Concentrated Sulfuric Acid and Leaching, JOM. 72 (2020) 816–822. https://doi.org/10.1007/s11837-019-03931-9.

[24] Mineral commodity summaries 2021, Reston, VA, 2021. https://doi.org/10.3133/mcs2021.

[25] N. Roxhed, P. Griss, G. Stemme, A method for tapered deep reactive ion etching using a modified Bosch process, J. Micromechanics Microengineering. 17 (2007) 1087–1092. https://doi.org/10.1088/0960-1317/17/5/031.

[26] K. Jung, W. Song, H.W. Lim, C.S. Lee, Parameter study for silicon grass formation in Bosch process, J. Vac. Sci. Technol. B. 28 (2010) 143–148. https://doi.org/10.1116/1.3280131.

[27] G.D. Demin, N.A. Djuzhev, N.A. Filippov, P.Y. Glagolev, I.D. Evsikov, N.N. Patyukov, Comprehensive analysis of field-electron emission properties of nanosized silicon blade-type and needle-type field emitters, J. Vac. Sci. Technol. B. 37 (2019) 22903. https://doi.org/10.1116/1.5068688.

[28] A. Haupt, Organic and Inorganic Fluorine Chemistry: Methods and Applications, De Gruyter, 2021. https://doi.org/doi:10.1515/9783110659337.

[29] P. Walker, W.H. Tarn, CRC Handbook of Metal Etchants, 1990. https://doi.org/10.1201/9781439822531.

[30] S. Middya, G.S. Kaminski Schierle, G.G. Malliaras, V.F. Curto, Chapter 8 - Lithography and electrodes, in: S. Das, S.B.T.-C.S.S. for M.D. and T.F.D.A. Dhara (Eds.), Elsevier, 2021: pp. 277–307. https://doi.org/https://doi.org/10.1016/B978-0-12-819718-9.00005-4.

[31] K.R. Williams, K. Gupta, M. Wasilik, Etch rates for micromachining processing-Part II, J. Microelectromechanical Syst. 12 (2003) 761–778. https://doi.org/10.1109/JMEMS.2003.820936.

[32] R. Löffler, M. Fleischer, D.P. Kern, An anisotropic dry etch process with fluorine chemistry to create well-defined titanium surfaces for biomedical studies, Microelectron. Eng. 97 (2012) 361–364. https://doi.org/10.1016/j.mee.2012.05.039.

[33] C. Cardinaud, Fluorine-based plasmas: Main features and application in micro-and nanotechnology and in surface treatment, Comptes Rendus Chim. 21 (2018) 723–739. https://doi.org/10.1016/j.crci.2018.01.009.

[34] K. Kim, A. Razzaq, S. Sorcar, Y. Park, C.A. Grimes, S.-I. In, Hybrid mesoporous $Cu_2ZnSnS_4$ (CZTS)–$TiO_2$ photocatalyst for efficient photocatalytic conversion of $CO_2$ into $CH_4$ under solar irradiation, RSC Adv. 6 (2016) 38964–38971. https://doi.org/10.1039/C6RA02763F.

[35] P. Pal, K. Sato, A comprehensive review on convex and concave corners in silicon bulk micromachining based on anisotropic wet chemical etching, Micro Nano Syst. Lett. 3 (2015) 6. https://doi.org/10.1186/s40486-015-0012-4.

[36] T. Ma, J. Wang, D. Li, Curvature-Modulated Si Spherical Cap-Like Structure Fabricated by Multistep Ring Edge Etching, Micromachines . 11 (2020). https://doi.org/10.3390/mi11080764.

[37] N. Asai, H. Ohta, F. Horikiri, Y. Narita, T. Yoshida, T. Mishima, Impact of damage-free wet etching process on fabrication of high breakdown voltage GaN p–n junction diodes, Jpn. J. Appl. Phys. 58 (2019) SCCD05. https://doi.org/10.7567/1347-4065/ab0401.

[38] L. Ni, M.P. de Boer, Self-Actuating Isothermal Nanomechanical Test Platform for Tensile Creep Measurement of Freestanding Thin Films, J. Microelectromechanical Syst. 31 (2022) 167–175. https://doi.org/10.1109/JMEMS.2021.3131153.

[39] L. Ni, R.M. Pocratsky, M.P. de Boer, Demonstration of tantalum as a structural material for MEMS thermal actuators, Microsyst. Nanoeng. 7 (2021) 6. https://doi.org/10.1038/s41378-020-00232-z.

[40] R. Kallio, P. Tanskanen, S. Luukkanen, Magnetic Preconcentration and Process Mineralogical Study of the Kiviniemi Sc-Enriched Ferrodiorite, Minerals. 11 (2021) 966. https://doi.org/10.3390/min11090966.

[41] T.J. Boyle, J.M. Sears, M.L. Neville, T.M. Alam, V.G. Young, Structural Properties of the Acidification Products of Scandium Hydroxy Chloride Hydrate, Inorg. Chem. 54 (2015) 11831–11841. https://doi.org/10.1021/acs.inorgchem.5b02030.

[42] H. Jansen, H. Gardeniers, M. de Boer, M. Elwenspoek, J. Fluitman, A survey on the reactive ion etching of silicon in microtechnology, J. Micromechanics Microengineering. 6 (1996) 14–28. https://doi.org/10.1088/0960-1317/6/1/002.

[43] H. Jansen, M. de Boer, J. Burger, R. Legtenberg, M. Elwenspoek, The black silicon method II:The effect of mask material and loading on the reactive ion etching of deep silicon trenches, Microelectron. Eng. 27 (1995) 475–480. https://doi.org/10.1016/0167-9317(94)00149-O.

[44] C.M. Silvestre, V. Nguyen, H. Jansen, O. Hansen, Deep reactive ion etching of 'grass-free' widely-spaced periodic 2D arrays, using sacrificial structures, Microelectron. Eng. 223 (2020) 111228. https://doi.org/10.1016/j.mee.2020.111228.

[45] M.J. Biercuk, D.J. Monsma, C.M. Marcus, J.S. Becker, R.G. Gordon, Low-temperature atomic-layer-deposition lift-off method for microelectronic and nanoelectronic applications, Appl. Phys. Lett. 83 (2003) 2405–2407. https://doi.org/10.1063/1.1612904.